\documentclass[aps,preprintnumbers,amsmath,amssymb,twocolumn,superscriptaddress,showpacs,floatfix]{revtex4}

\usepackage{graphicx}
\usepackage{dcolumn}
\usepackage{bm}

\begin{document}
\title{Azimuthal and single spin asymmetry in deep-inelastic lepton-nucleon scattering}
\author{Zuo-tang Liang$^{1,2}$ and Xin-Nian Wang$^{2,}$}
\affiliation{Department of Physics, Shandong University, Jinan, Shandong 250100, China\\
$^2$Nuclear Science Division, MS 70R0319,
Lawrence Berkeley National Laboratory, Berkeley, California 94720}

\date{\today}

\preprint {LBNL-61449}

\begin{abstract}
We derive a general framework for describing semi-inclusive 
deep-inelastic lepton-nucleon scattering in terms of the 
unintegrated parton distributions and other 
higher twist parton correlations.
Such a framework provides a consistent 
approach to the calculation of inclusive and semi-inclusive 
cross sections including higher twist effects.
As an example, we calculate the azimuthal asymmetries to the order 
of $1/Q$ in semi-inclusive process with transversely polarized target. 
A non-vanishing single-spin asymmetry in the ``triggered inclusive process'' 
is predicted to be $1/Q$ suppressed with a part of the coefficient 
related to a moment of the Sivers function.
\end{abstract}


\maketitle

\section{Introduction}

Many interesting phenomena have been observed
\cite{Aubert:1983cz,Arneodo:1986cf,Adams:1993hs,Breitweg:2000qh,
Chekanov:2002sz,Airapetian:2004tw,Alexakhin:2005iw,Webb:2005cd} 
in semi-inclusive deep-inelastic lepton-nucleon scattering (SIDIS), 
in particular the azimuthal asymmetries in the momentum distribution 
of the final hadrons and their spin dependence 
\cite{Aubert:1983cz,Arneodo:1986cf,Adams:1993hs,Breitweg:2000qh,
Chekanov:2002sz,Airapetian:2004tw,Alexakhin:2005iw,Webb:2005cd,
Georgi:1977tv,Cahn:1978se,Berger:1979kz,Liang:1993re,Oganesian:1997jq,Chay:1997qy,
Collins:1992kk,Mulders:1995dh,Boer:1997nt,Boer:1999uu,Ji:2002aa,Belitsky:2002sm,Ji:2004wu,Ji:2005nu,Anselmino:2005ea}.
Since most of the studies involve hadrons with transverse momenta $p_\perp \sim 1$ GeV/c, 
the intrinsic parton transverse momenta (denoted by $k_\perp$) 
and multiple parton scattering 
become critical in the perturbative QCD (pQCD) approach.

The effects of intrinsic parton transverse momenta 
and multiple parton scattering are normally higher-twist.
The higher twist effects have been studied extensively 
in inclusive DIS or lepton-pair production in 
in hadron-hadron or hadron-nucleus
collisions in the past. 
An elegant and practical framework in terms of collinear expansion
has been developed and applied to 
these processes \cite{Ellis:1982wd,Qiu:1990xx,Guo:2000nz}.
A factorized form for the cross section 
is obtained as a convolution of the calculable hard parts   
with the universal ($k_\perp$-integrated) 
parton distributions and correlation functions 
(hereafter referred generally as parton correlation functions) 
that can be measured in different reactions.
The framework is important not only because 
it provides a practical way of studying higher twist effects 
but also because it leads to the definitions of 
the parton correlations in a gauge invariant form.

The gauge invariant parton correlations contain contributions 
from the initial and final state interactions with soft gluons,
which, when extended to include transverse momenta, 
lead in particular to the single-spin asymmetry in 
the transversely polarized case \cite{Ji:2002aa,Belitsky:2002sm,Ji:2004wu}.
The asymmetry can manifest itself in SIDIS, hence 
the framework has been generalized to study the 
azimuthal asymmetries in SIDIS
without explicit derivations \cite{Mulders:1995dh,Anselmino:2005ea}.  
It is not clear whether/how the collinear expansion can be made in SIDIS. 
Many of the generalized formulas used in the literature are not proved  
and some of them are in fact even erroneous, in particular 
when higher twist effects are involved. 

In this paper, we will clarify the situation by making an 
explicit derivation that generalizes the factorization 
to the transverse-momentum-dependent SIDIS.
We show that the collinear expansion can also be made in SIDIS and 
such a derivation leads to a general framework 
for describing SIDIS including higher twist effects. 
We present the results and show that they have 
a number of remarkable properties that are 
often neglected in the literature.
The framework in this study provides a consistent and 
systematic approach to the pQCD study of SIDIS beyond the leading twist. 
As an example, we present the calculation of 
the differential cross section for SIDIS 
with transversely polarized target to the order $1/Q$. 
We show in particular that the results imply the 
existence of a new single-spin asymmetry 
in the ``triggered inclusive process'' 
which can easily be tested experimentally.

The paper is organized as follows. 
After this introduction, we will make  
a short review of the key ingredients of the collinear expansion technique 
and its application to inclusive DIS, 
then show how it can be applied to SIDIS in Sec.~II. 
In Sec. III, as an example of the application of the formulas obtained 
in Sec. II, we will present the differential cross section and 
azimuthal asymmetries in SIDIS using unpolarized electron and 
transversely polarized nucleon. 
Finally, we make a short summary in Sec.~IV. 

\section{Collinear expansion in SIDIS}

\subsection{Collinear expansion in inclusive DIS}

A general framework for studying inclusive DIS including 
higher twist contributions has been developed in Refs.\cite{Ellis:1982wd,Qiu:1990xx}. 
We review the key ingredients here in order to show how they can be extended to SIDIS.  
We consider the inclusive DIS $e^-p\to e^-X$, and the differential cross section is given by,
\begin{equation}
d\sigma=\frac{e^4}{sQ^4}L^{\mu\nu}(l,l')W_{\mu\nu}(q,p,S)
\frac{d^3l'}{(2\pi)^3 2E_{l'}},
\end{equation}
where $l$ and $l'$ are respectively the four momenta of the incoming and outgoing leptons, 
$p$ and $S$ are the four momentum  and the spin of the incoming proton, 
$q$ is the four momentum transfer. 
We neglect the masses and use the light-cone coordinates.
The unit vectors are taken as,  
$\bar n=(1,0,0,0)$, $n=(0,1,0,0)$, $n_{\perp 1}=(0,0,1,0)$,
$n_{\perp 2}=(0,0,0,1)$. 
We work in the center of mass frame of the $\gamma^*p$-system, 
and chose the coordinate system in the way so that, 
$p=p^+\bar n$, $q=-x_Bp+nQ^2/(2x_Bp^+)$, and 
$l_\perp=|\vec l_\perp|n_{\perp 1}$,  
where $x_B=Q^2/2p\cdot q$ is the Bjorken-x and $y=p\cdot q/p\cdot l$.
The leptonic tensor $L^{\mu\nu}$ 
is defined as usual and is given by,
\begin{equation}
L^{\mu\nu}(l,l')
=4[l^\mu{l'}^\nu+l^\nu{l'}^\mu-(l\cdot l')g^{\mu\nu}],
\end{equation}
The hadronic tensor $W_{\mu\nu}$ is defined as, 
\begin{widetext}
\begin{equation}
W_{\mu\nu}(q,p,S)=
\frac{1}{2\pi}\sum_X
\langle p,S| J_\mu(0)|X\rangle \langle X| J_\nu(0)|p,S\rangle (2\pi)^4\delta^4(p+q-p_X),
\end{equation}

\begin{figure}[htbp]
\includegraphics[width=12cm]{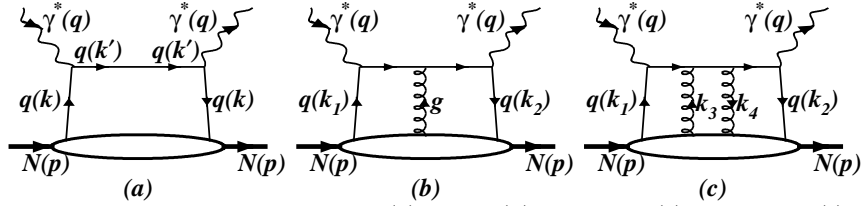}
\caption{Feynman diagrams for the cases 
with exchange of (a) $j=0$, (b) $j=1$ and (c) $j=2$ gluon(s). 
The gluon momentum in (b) is $k=k_1-k_2$, 
those in (c) are $k_3=k-k_1$ and $k_4=k-k_2$.}
\label{fig1}
\end{figure}

We consider final state interaction in pQCD 
so that we have the contributions from the type of diagrams shown in Fig.1.
The hadronic tensor $W_{\mu\nu}$ should be written as a sum of the  
contributions from all the diagrams, i.e., 
$W_{\mu\nu}=\sum_j W_{\mu\nu}^{(j)}$, 
where $j$ denotes the number of soft gluons.
At the lowest order in pQCD, we have, 
\begin{eqnarray}
W_{\mu\nu}^{(0)}(q,p,S)&=&\frac{1}{2\pi}
\int\frac{d^4k}{(2\pi)^4}
{\rm Tr}[\hat H_{\mu\nu}^{(0)}(k,q)\hat \phi^{(0)}(k,p,S)],\\
\hat H_{\mu\nu}^{(0)}(k,q)&=&
\gamma_\mu(\slash{\hspace{-5pt}k}+\slash{\hspace{-5pt}q})\gamma_\nu
 (2\pi)\delta_+((k+q)^2),
\end{eqnarray}
where $\delta_+$ means that only the positive solution is taken.
Similarly, corresponding to Figs.~1(b) and (c), we have, 
\begin{eqnarray}
W_{\mu\nu}^{(1)}(q,p,S)&=&
\frac{1}{2\pi}\int\frac{d^4k_1}{(2\pi)^4}\frac{d^4k_2}{(2\pi)^4} 
{\rm Tr}[\hat H_{\mu\nu}^{(1)\rho}(k_1,k_2,q) 
\hat \phi^{(1)}_\rho(k_1,k_2,p,S)], \\ 
W_{\mu\nu}^{(2)}(q,p,S)&=&
\frac{1}{2\pi}\int\frac{d^4k_1}{(2\pi)^4}\frac{d^4k_2}{(2\pi)^4}\frac{d^4k}{(2\pi)^4}
{\rm Tr}[\hat H_{\mu\nu}^{(2)\rho\sigma}(k_1,k_2,k,q) 
\hat \phi^{(2)}_{\rho\sigma}(k_1,k_2,k,p,S)],
\end{eqnarray}
where $\hat H_{\mu\nu}^{(1)\rho}(k_1,k_2,q)=\sum_{c=L,R}\hat H_{\mu\nu}^{(1,c)\rho}(k_1,k_2,q)$, 
$\hat H_{\mu\nu}^{(2)\rho}(k_1,k_2,k,q)=\sum_{c=L,M,R}\hat H_{\mu\nu}^{(2,c)\rho}(k_1,k_2,k,q)$, 
and $c$ denotes the different cuts in the diagrams.
These hard parts can all be read from the diagram and are given by,
\begin{eqnarray}
\hat H_{\mu\nu}^{(1,L)\rho}(k_1,k_2,q)&=&
\gamma_\mu(\slash{\hspace{-5pt}k_1}+\slash{\hspace{-5pt}q})\gamma^\rho
\frac{\slash{\hspace{-5pt}k_2}+\slash{\hspace{-5pt}q}}{(k_2+q)^2-i\epsilon}
\gamma_\nu (2\pi)\delta_+((k_1+q)^2),\\
\hat H_{\mu\nu}^{(2,L)\rho\sigma}(k_1,k_2,k,q)&=&
\gamma_\mu (\slash{\hspace{-5pt}k_1}+\slash{\hspace{-5pt}q})\gamma^\rho
\frac{\slash{\hspace{-5pt}k}+\slash{\hspace{-5pt}q}}
{(k+q)^2-i\epsilon}\gamma^\sigma
\frac{\slash{\hspace{-5pt}k_2}+\slash{\hspace{-5pt}q}}
{(k_2+q)^2-i\epsilon}
\gamma_\nu (2\pi)\delta_+((k_1+q)^2),
\end{eqnarray}
and so on. 
The structure of proton is contained only in the matrix elements $\hat\phi$'s that are defined as,
\begin{eqnarray}
\hat\phi^{(0)}(k,p,S) &\equiv & 
\int d^4ze^{ikz}
\langle p,S|\bar\psi(0)\psi(z)|p,S\rangle, \\ 
\hat\phi^{(1)}_\rho(k_1,k_2,p,S) &\equiv & 
\int d^4yd^4ze^{ik_1y+ik_2(z-y)}
\langle p,S|\bar\psi(0)gA_\rho(y)\psi(z)|p,S\rangle, \\
\hat\phi^{(2)}_{\rho\sigma}(k_1,k_2,k,p,S) &\equiv & 
\int d^4yd^4y'd^4ze^{ik_1y+ik(y'-y)+ik_2(z-y')}
\langle p,S|\bar\psi(0)gA_\rho(y)gA_\sigma(y')\psi(z)|p,S\rangle.
\end{eqnarray}

We note that neither of the $\hat\phi$'s defined in this way is gauge invariant.
To organize the above results in terms of gauge invariant parton correlations, 
we need to invoke the collinear expansion procedure. 
This procedure has been developed in Refs.\cite{Ellis:1982wd,Qiu:1990xx}, 
and is carried out in the following steps. 

(1) we make a Taylor expansion of the hard parts around $k=xp$, e.g., 
\begin{equation}
\hat H^{(0)}_{\mu\nu}(k,q)=
\hat H^{(0)}_{\mu\nu}(x)+
\frac{\partial\hat H^{(0)}_{\mu\nu}(x)}
{\partial k_\rho}\omega_\rho^{\ \rho'}k_{\rho'}+
\frac{1}{2}
\frac{\partial^2\hat H^{(0)}_{\mu\nu}(x)}{\partial k_\rho \partial k_\sigma}
\omega_\rho^{\ \rho'}\omega_\sigma^{\ \sigma'}k_{\rho'}k_{\omega'}+ ...
\end{equation}
where $x=k^+/p^+$, 
$\hat H^{(0)}_{\mu\nu}(x)\equiv \hat H^{(0)}_{\mu\nu}(k,q)|_{k=xp}$, 
$\partial\hat H^{(0)}_{\mu\nu}(x)/\partial k_\rho
\equiv \partial\hat H^{(0)}_{\mu\nu}(k,q)/\partial k_\rho|_{k=xp}$ and so on; 
$\omega_{\rho}^{\ \rho'}\equiv g_\rho^{\ \rho'}-\bar n_\rho n^{\rho'}$ 
is a projection operator 
so that $\omega_{\rho}^{\ \rho'}k_{\rho'}=(k-xp)_\rho$.

(2) we decompose the gluon field $A_\rho$ into
the longitudinal and transverse components, i.e., 
\begin{equation}
A_\rho(y)=\omega_\rho^{\ \rho'}A_{\rho'}(y)+p_\rho n\cdot A(y)/n\cdot p. 
\end{equation}

(3) we use the generalized Ward identities to relate 
the derivative of a hard part to that of a higher order, 
and the product of $p$ with a hard part to that of a lower order, 
{\it i.e.}, 
\begin{eqnarray}
\label{eq:wd1}
&& \frac{\partial\hat H^{(0)}_{\mu\nu}(x)}{\partial k_\rho} =-\hat H^{(1)\rho}_{\mu\nu}(x,x),
\hspace{1cm} 
\frac{1}{2}\frac{\partial^2\hat H^{(0)}_{\mu\nu}(x)}{\partial k_\rho\partial k_\sigma} =\hat H^{(2)\rho\sigma}_{\mu\nu}(x,x,x),\\
\label{eq:wd2}
&& \frac{\partial\hat H^{(1)\rho}_{\mu\nu}(x_1,x_2)}{\partial k_{2\sigma}}= -\hat H^{(2)\rho\sigma}_{\mu\nu}(x_1,x_2,x_2), 
\hspace{1cm} 
\frac{\partial\hat H^{(1)\rho}_{\mu\nu}(x_1,x_2)}{\partial k_{1\sigma}}= -\hat H^{(2)\sigma\rho}_{\mu\nu}(x_1,x_1,x_2), \\
\label{eq:wd3}
&& p_\rho\hat H^{(1)\rho}_{\mu\nu}(x_1,x_2)=
\frac{\hat H^{(0)}_{\mu\nu}(x_1)}{x_2-x_B-i\epsilon}+\frac{\hat H^{(0)}_{\mu\nu}(x_2)}{x_1-x_B+i\epsilon}, \\
\label{eq:wd4}
&& p_\rho\hat H^{(2)\rho\sigma}_{\mu\nu}(x_1,x_2,x)=
\frac{\hat H^{(1)\sigma}_{\mu\nu}(x_1,x_2)}{x-x_1-i\epsilon}+\frac{\hat H^{(1)\sigma}_{\mu\nu}(x,x_2)}{x_1-x+i\epsilon}, \\
\label{eq:wd5}
&& p_\sigma\hat H^{(2)\rho\sigma}_{\mu\nu}(x_1,x_2,x)=
\frac{\hat H^{(1)\rho}_{\mu\nu}(x_1,x)}{x_2-x-i\epsilon}+\frac{\hat H^{(1)\rho}_{\mu\nu}(x_1,x_2)}{x-x_2+i\epsilon}, 
\end{eqnarray}

(4) we rearrange all the terms by adding the contributions 
with the same hard part together. 
In this way, we  obtain that 
$W_{\mu\nu}(q,p,S)=\sum_{j=0,1,2}\tilde W_{\mu\nu}^{(j)}(q,p,S)$, 
and,
\begin{eqnarray}
\label{eq:Winc1}
&&\tilde W_{\mu\nu}^{(0)}(q,p,S)=
\frac{1}{2\pi}\int dx {\rm Tr}[\hat H_{\mu\nu}^{(0)}(x)\ \hat \Phi^{(0)}(x,p,S)]; \\
\label{eq:Winc2}
&&\tilde W_{\mu\nu}^{(1)}(q,p,S)=\frac{1}{2\pi}\int dx_1dx_2
{\rm Tr}[\hat H_{\mu\nu}^{(1)\rho}(x_1,x_2) 
\omega_\rho^{\ \rho'} \hat \Phi^{(1)}_{\rho'}(x_1,x_2,p,S)];\\
\label{eq:Winc3}
&&\tilde W_{\mu\nu}^{(2)}(q,p,S)=\frac{1}{2\pi} \int dx_1dx_2dx 
{\rm Tr}[\hat H_{\mu\nu}^{(2)\rho\sigma}(x_1,x_2,x)  
\omega_\rho^{\ \rho'}\omega_\sigma^{\ \sigma'}
\hat\Phi^{(2)}_{\rho'\sigma'}(x_1,x_2,x,p,S)], 
\end{eqnarray}
where the matrix elements $\hat\Phi$'s have contributions 
from all the three diagrams in Fig.~1, and are given by,
\begin{eqnarray}
\label{eq:Phi0}
&&\hat\Phi^{(0)}(x,p,S)=\int dz^- e^{ixp^+z^-}
\langle p,S|\bar\psi(0){\cal L}(0,z^-)
\psi(z^-)|p,S\rangle, \\ 
\label{eq:Phi1}
&&\hat\Phi^{(1)}_\rho(x_1,x_2,p,S)
=\int dy^-dz^-e^{ix_1p^+y^-+ix_2p^+(z^--y^-)}
\langle p,S|\bar\psi(0) {\cal L}(0,y^-)D_\rho(y^-){\cal L}(y^-,z^-)\psi(z^-)|p,S\rangle, \\
\label{eq:Phi2}
&& \hat\Phi^{(2)}_{\rho\sigma}(x_1,x_2,x,p,S)=\int dy^-d{y'}^-dz^-
e^{ix_1p^+y^-+ixp^+({y'}^--y^-)+ix_2p^+(z^--{y'}^-)} \nonumber\\
&& \phantom{XXXXXXXXXXXX} \times 
\langle p,S|\bar\psi(0){\cal L}(0,y^-)
D_\rho(y^-) {\cal L}(y^-,{y'}^-)D_\sigma({y'}^-){\cal L}({y'}^-,z^-)\psi(z^-)|p,S\rangle, 
\end{eqnarray}
where $D_\rho(y)=-i\partial_\rho+gA_\rho(y)$ is the covariant derivative.
${\cal L}(y,z)$ is the so-called gauge link, which is obtained in the derivation and is given by, 
\begin{equation}
{\cal L}(y^-,z^-)=
1+ig\int_{y^-}^{z^-}d\xi^-A^+(0,\xi^-,\vec 0_\perp)
+(ig)^2 \int_{y^-}^{z^-}d\xi^- \int^{\xi^-}_{y^-}d\eta^- 
A^+(0,\xi^-,\vec 0_\perp)A^+(0,\eta^-,\vec 0_\perp),
\end{equation}
Including higher order contributions in $g$, the gauge link ${\cal {L}}$ 
is given by, 
\begin{equation}
{\cal L}(y^-,z^-)
=Pe^{ig\int_{y^-}^{z^-}d\xi^- A^+(0,\xi^-,\vec 0_\perp)},
\end{equation}
which is the well-known path integral representation of the gauge link 
without transverse displacement.
These $\hat\Phi$'s given by Eqs.(\ref{eq:Phi0}-\ref{eq:Phi2}) are now gauge invariant. 
They lead to the well-known gauge invariant definitions of the parton correlation functions 
used in this field.\cite{Ellis:1982wd,Qiu:1990xx,Guo:2000nz}

\subsection{Generalization to SIDIS}

Now we apply the same procedure to SIDIS. 
We show why and how the collinear expansion can also be made for SIDIS.
To show the main idea, we start with a simple case 
$e^-p^\uparrow\to e^-qX$, i.e., 
we do not consider the fragmentation. 
This is equivalent to consider jet production. 
The differential cross section for 
$e^-p^\uparrow\to e^-qX$ can be written as, 
\begin{equation}
d\sigma=\frac{\alpha_{em}^2e_q^2}{sQ^4}L^{\mu\nu}(l,l')W_{\mu\nu}^{(si)}(q,p,S,k')
\frac{d^3l'd^3k'}{(2\pi)^4E_{l'}E_{k'}},
\end{equation}
where $k'$ is the 4-momentum of the outgoing quark; 
and the hadronic tensor $W_{\mu\nu}^{(si)}$ for SIDIS   
is defined as, 
\begin{equation}
W_{\mu\nu}^{(si)}(q,p,S,k')=
\frac{1}{2\pi}\sum_X
\langle p,S| J_\mu(0)|k',X\rangle \langle k',X| J_\nu(0)|p,S\rangle 
(2\pi)^4\delta^4(p+q-k'-p_X),
\end{equation}
where the superscript $(si)$ denotes that it is for SIDIS.

We consider the same final state interaction in pQCD as one did for inclusive DIS 
so that we have the contributions from the diagrams shown in Fig.1.
The hadronic tensor $W_{\mu\nu}^{(si)}$ is written as a sum of the  
contributions from all the diagrams, i.e., 
$W_{\mu\nu}^{(si)}=\sum_j W_{\mu\nu}^{(j,si)}$,  
\begin{eqnarray}
&&W_{\mu\nu}^{(0,si)}(q,p,S,k')=\frac{1}{2\pi}
\int\frac{d^4k}{(2\pi)^4}
{\rm Tr}[\hat H_{\mu\nu}^{(0,si)}(k,k',q)\hat \phi^{(0)}(k,p,S)], \\
&&W_{\mu\nu}^{(1,si)}(q,p,S,k')
=\frac{1}{2\pi}\int\frac{d^4k_1}{(2\pi)^4}\frac{d^4k_2}{(2\pi)^4} 
\sum_{c=L,R} {\rm Tr}[\hat H_{\mu\nu}^{(1,c,si)\rho}(k_1,k_2) 
\hat \phi^{(1)}_\rho(k_1,k_2,p,S)]; \\
&&W_{\mu\nu}^{(2,si)}(q,p,S,k')
=\frac{1}{2\pi}\int\frac{d^4k_1}{(2\pi)^4}\frac{d^4k_2}{(2\pi)^4} \frac{d^4k}{(2\pi)^4} 
\sum_{c=L,M,R} {\rm Tr}[\hat H_{\mu\nu}^{(2,c,si)\rho\sigma}(k_1,k_2,k) 
\hat \phi^{(2)}_{\rho\sigma}(k_1,k_2,k,p,S)].\  
\end{eqnarray}
where $\hat H_{\mu\nu}^{(j,c,si)}$'s denote the hard parts for the 
semi-inclusive process and are given by,
\begin{eqnarray} 
&&\hat H_{\mu\nu}^{(0,si)}(k,k',q)=
\gamma_\mu(\slash{\hspace{-5pt}k}+\slash{\hspace{-5pt}q})\gamma_\nu
(2\pi)^4\delta^4(k'-k-q),\\
&&\hat H_{\mu\nu}^{(1,L,si)\rho}(k_1,k_2,q)=
\gamma_\mu(\slash{\hspace{-5pt}k_1}+\slash{\hspace{-5pt}q})\gamma^\rho
\frac{\slash{\hspace{-5pt}k_2}+\slash{\hspace{-5pt}q}}{(k_2+q)^2-i\epsilon}
\gamma_\nu (2\pi)^4\delta^4(k'-k_1-q),\\
&&\hat H_{\mu\nu}^{(2,L,si)\rho\sigma}(k_1,k_2,k,q)=
\gamma_\mu (\slash{\hspace{-5pt}k_1}+\slash{\hspace{-5pt}q})\gamma^\rho
\frac{\slash{\hspace{-5pt}k}+\slash{\hspace{-5pt}q}}
{(k+q)^2-i\epsilon}\gamma^\sigma
\frac{\slash{\hspace{-5pt}k_2}+\slash{\hspace{-5pt}q}}
{(k_2+q)^2-i\epsilon}\gamma_\nu 
(2\pi)^4\delta^4(k'-k_1-q),
\end{eqnarray}
We compare the $W_{\mu\nu}^{(j,si)}$'s with their counterparts 
$W_{\mu\nu}^{(j)}$ for the inclusive reaction, 
where an integration over $d^3k'$ is carried out, 
and we see that 
they differ from each other only in the hard parts.
If we would now apply the same procedure as we did for inclusive DIS 
to make the collinear expansion, i.e., we make a Taylor expansion 
for the semi-inclusive hard parts $H_{\mu\nu}^{(j,si)}$ directly, 
we would find out that there are no such relations valid 
for the semi-inclusive hard parts 
as given in Eqs.(\ref{eq:wd1}-\ref{eq:wd5}) for the inclusive hard parts.
We would not be able to reach similar results as given in Eqs.(\ref{eq:Winc1}-\ref{eq:Winc3}).

This difficulty can be avoided by using the identity, 
\begin{equation} 
\delta^4(k'-k-q)=2E_{k'}\delta_+((q+k)^2)\delta^3(\vec k'-\vec k-\vec q), 
\end{equation}
to rewrite the hard parts for the semi-inclusive processes. 
Using this identity, we have,
\begin{eqnarray}
&&\hat H_{\mu\nu}^{(0,si)}(k,k',q)=\hat H_{\mu\nu}^{(0)}(k,q)K(k,k',q), \\
&&K(k',k,q)\equiv 2E_{k'}(2\pi)^3\delta^3(\vec k'-\vec k-\vec q).
\end{eqnarray}
We see that the difference between $\hat H_{\mu\nu}^{(0,si)}(k,k',q)$
and its counterpart $\hat H_{\mu\nu}^{(0)}(k,q)$ is only 
a multiplicative factor $K(k',k,q)$. 
Correspondingly, for the hadronic tensor, we have, 
\begin{equation}
W_{\mu\nu}^{(0,si)}(q,p,S,k')
=\frac{1}{2\pi}\int\frac{d^4k}{(2\pi)^4} K(k',k,q)
{\rm Tr}[\hat H_{\mu\nu}^{(0)}(k,q)\hat \phi^{(0)}_\rho(k,p,S)]. 
\end{equation}
This is also true for higher number of gluon exchanges, 
i.e., for higher $j$ as shown in Figs. 1(b) and 1(c), 
but the kinematic factor $K$ is different for different cuts.  
E.g., for $j=1$ and $2$, corresponding to Figs.~1(b) and 1(c), we have, 
\begin{eqnarray}
&& \hat H_{\mu\nu}^{(1,c,si)\rho}(k_1,k_2,k',q)=\hat H_{\mu\nu}^{(1,c)\rho}(k_1,k_2,q)K(k_c,k',q), \\
&& \hat H_{\mu\nu}^{(2,c,si)\rho\sigma}(k_1,k_2,k,k',q)=\hat H_{\mu\nu}^{(2,c)\rho\sigma}(k_1,k_2,k,q)K(k_c,k',q).
\end{eqnarray}
where $c=L, R$ or $M$ and $k_L=k_1$, $k_R=k_2$, $k_M=k$.  
Correspondingly, we have, 
\begin{eqnarray}
&&W_{\mu\nu}^{(1,si)}(q,p,S,k')=
\frac{1}{2\pi}\int\frac{d^4k_1}{(2\pi)^4}\frac{d^4k_2}{(2\pi)^4}\sum_{c=L,R}K(k',k_c,q)
{\rm Tr}[\hat H_{\mu\nu}^{(1,c)\rho}(k_1,k_2,q) 
\hat \phi^{(1)}_\rho(k_1,k_2,p)]; \\
&&W_{\mu\nu}^{(2,si)}(q,p,S,k')=
\frac{1}{2\pi}\int\frac{d^4k_1}{(2\pi)^4}\frac{d^4k_2}{(2\pi)^4}\frac{d^4k}{(2\pi)^4}
\sum_{c=L,R,M} K(k',k_c,q)
{\rm Tr}[\hat H_{\mu\nu}^{(2,c)\rho\sigma}(k_1,k_2,k,q) 
\hat \phi^{(2)}_{\rho\sigma}(k_1,k_2,k,p,S),
\end{eqnarray}

This feature is very important since a multiplicative factor 
does not influence the collinear expansion that are needed to make 
the definition of the parton correlations gauge invariant. 
It implies that we can apply the same collinear expansion procedure as that for inclusive processes.
More precisely, we do in exactly the same way as we did for inclusive DIS. 
We go through the steps (1) to (4) given in section A and 
in step (1) we make a Taylor expansion of the corresponding inclusive hard parts $\hat H$'s 
but {\it not} the semi-inclusive $\hat H^{(si)}$'s around $k=xp$. 
In this way, in step (3), we need similar relations between the derivatives
of the inclusive hard parts and those of a higher order.
The only differnce is that we need to generalize them to the hard parts with different cuts separately 
since the kinetic factor $K$ depends on the cut. 
This generalization can be derived and the results obtained are similar, e.g., 
\begin{eqnarray}
&& \frac{\partial\hat H^{(0)}_{\mu\nu}(x)}{\partial k_\rho} =
  -\hat H^{(1,L)\rho}_{\mu\nu}(x,x)-\hat H^{(1,R)\rho}_{\mu\nu}(x,x),\\
&& \frac{\partial\hat H^{(1,L)\rho}_{\mu\nu}(x_1,x_2)}{\partial k_{2\sigma}}=
 -\hat H^{(2,L)\rho\sigma}_{\mu\nu}(x_1,x_2,x_2), \\
&& \frac{\partial\hat H^{(1,L)\rho}_{\mu\nu}(x_1,x_2)}{\partial k_{1\sigma}}=
 -\hat H^{(2,L)\rho\sigma}_{\mu\nu}(x_1,x_2,x_1)
 -\hat H^{(2,M)\rho\sigma}_{\mu\nu}(x_1,x_2,x_1),\\
&& p_\rho\hat H^{(1,L)\rho}_{\mu\nu}(x_1,x_2)=
 \frac{\hat H^{(0)}_{\mu\nu}(x_1)}{x_2-x_1-i\epsilon}, \\
&& p_\rho p_\sigma\hat H^{(2,L)\rho\sigma}_{\mu\nu}(x_1,x_2,x)=
 \frac{\hat H^{(0)}_{\mu\nu}(x_1)}{(x_2-x_1-i\epsilon)(x-x_1-i\epsilon)}, \\
&& p_\rho\hat H^{(2,R)\rho\sigma}_{\mu\nu}(x_1,x_2,x)=
\frac{\hat H^{(1,M)\sigma}_{\mu\nu}(x_1,x_2)-
 \hat H^{(1,M)\sigma}_{\mu\nu}(x,x_2)}{x-x_1-i\epsilon}. 
\end{eqnarray}
Finally, we obtain similar results as, 
$W_{\mu\nu}^{(si)}(q,p,S,k')=\sum_{j=0,1,2}\tilde W_{\mu\nu}^{(j,si)}(q,p,S,k')$, 
and,
\begin{eqnarray}
\label{eq:Wsi1}
&& \tilde W_{\mu\nu}^{(0,si)}
=\frac{1}{2\pi}
\int\frac{d^4k}{(2\pi)^4}K(k',k,q)
{\rm Tr}[\hat H_{\mu\nu}^{(0)}(x)\ \hat \Phi^{(0)}(k,p,S)];\\
\label{eq:Wsi2}
&&\tilde W_{\mu\nu}^{(1,si)}
=\frac{1}{2\pi}
\int\frac{d^4k_1}{(2\pi)^4}\frac{d^4k_2}{(2\pi)^4}\sum_{c=L,R}K(k',k_c,q) 
{\rm Tr}[\hat H_{\mu\nu}^{(1,c)\rho}(x_1,x_2) \omega_\rho^{\ \rho'} 
\hat \Phi^{(1)}_{\rho'}(k_1,k_2,p,S)];\\
\label{eq:Wsi3}
&&\tilde W_{\mu\nu}^{(2,si)}
=\frac{1}{2\pi}
\int\frac{d^4k_1}{(2\pi)^4}\frac{d^4k_2}{(2\pi)^4}\frac{d^4k}{(2\pi)^4}\sum_{c=L,R,M}
K(k',k_c,q){\rm Tr}[\hat H_{\mu\nu}^{(2,c)\rho\sigma}(x_1,x_2,x)  
\omega_\rho^{\ \rho'}\omega_\sigma^{\ \sigma'}
\hat\Phi^{(2)}_{\rho'\sigma'}(k_1,k_2,k,p,S)].\ 
\end{eqnarray}
Because of the existence of the factor $K$, 
we can not carry out the integration over $k^-$ and $\vec k_\perp$ as we did 
for inclusive cross section.
We have to keep the un-integrated gauge invariant matrix elements $\hat\Phi$'s. 
They have received the contributions from all the three diagrams in Fig.~1,
and are given by,
\begin{eqnarray}
\label{eq:Phiun1}
&&\hat\Phi^{(0)}(k,p,S)=\int d^4ze^{ikz}
\langle p,S|\bar\psi(0){\cal L}(0,z) \psi(z)|p,S\rangle, \\
\label{eq:Phiun2}
&&\hat\Phi^{(1)}_\rho(k_1,k_2,p,S)
=\int d^4yd^4ze^{ik_1y+ik_2(z-y)}
\langle p,S|\bar\psi(0) {\cal L}(0,y)D_\rho(y){\cal L}(y,z)\psi(z)|p,S\rangle,\\
\label{eq:Phiun3}
&&\hat\Phi^{(2)}_{\rho\sigma}(k_1,k_2,k,p,S)=\int d^4yd^4y'd^4z
e^{ik_1y+ik(y'-y)+ik_2(z-y')} \langle p,S|\bar\psi(0){\cal L}(0,y)
D_\rho(y) {\cal L}(y,y')D_\sigma(y'){\cal L}(y',z)\psi(z)|p,S\rangle,  \nonumber
\end{eqnarray}
where ${\cal L}(0,z)$ is the gauge link with transverse separation. 
${\cal L}(0,z)$ is obtained explicitly from the derivation 
and it is a product of two parts, {\it i.e.}, 
\begin{eqnarray} 
&&{\cal L}(0,z)={\cal L}^\dag (-\infty,0){\cal L}(-\infty,z), \\
&&{\cal L}(-\infty,z)
=1+ig\int_{-\infty}^{z^-}dy^-A^+(z^+,y^-,\vec z_\perp)
+(ig)^2 \int_{-\infty}^{z^-}dy^-\int_{-\infty}^{y^-}d{y'}^-
A^+(z^+,y^-,\vec z_\perp)A^+(z^+,{y'}^-,\vec z_\perp),
\end{eqnarray}
Including higher order contributions in $g$, we have,
\begin{equation}
{\cal L}(-\infty,z)=Pe^{ig\int_{-\infty}^{z^-}dy^-n\cdot A(z^+,y^-,\vec z_\perp)}.
\end{equation} 
which is the path integral expression 
in the case that there is a transverse 
displacement $\vec{z}_\perp$ \cite{ftQiu}. 

Eqs.(\ref{eq:Wsi1}-\ref{eq:Wsi3}) can be considered as a generalized factorization 
of transverse momentum dependent SIDIS as a result of the collinear expansion.
Consequently, there are two distinct properties in the factorized form:

(A) All the hard parts in the Eqs.~(\ref{eq:Wsi1}-\ref{eq:Wsi3}) are only
functions of the longitudinal parton momenta. 
All the information of transverse momenta is contained 
in the matrix elements $\hat\Phi$'s.

(B) The operator
$\omega_\rho^{\ \rho'}$ 
in $\tilde W_{\mu\nu}^{(1,si)}$ 
projects away the longitudinal components of the gauge fields 
that go into the gauge links.   

We want to point out that the final result 
for the cross section is {\it not} 
simply a convolution of the {\it transverse momentum dependent}
lepton-quark scattering cross section with the 
unintegrated (or $k_\perp$-dependent) quark distributions
as in some of the existing literature. 
Such a convolution can result in double-counting of the 
effects of transverse momenta.

\section{Differential cross section and azimuthal asymmetries for SIDIS}

The properties (A) and (B) mentioned above lead to 
a great simplification of $\tilde{W}^{(1,si)}_{\mu\nu}$.
It can be shown that
\begin{eqnarray}
&&H_{\mu\nu}^{(1,L)\rho}(x_1,x_2)\omega_\rho^{\ \rho'}
=\frac{\pi}{2q\cdot p}\delta(x_1-x_B)
\gamma_\mu\slash{\hspace{-5pt}n}\gamma^\rho\slash{\hspace{-5pt}\bar n}\gamma_\nu \omega_\rho^{\ \rho'} 
\equiv \hat H_{\mu\nu}^{(1)\rho}(x_1)\omega_\rho^{\ \rho'}, \\ 
&&H_{\mu\nu}^{(1,R)\rho}(x_1,x_2)\omega_\rho^{\ \rho'}
=\frac{\pi}{2q\cdot p}\delta(x_2-x_B)
\gamma_\mu\slash{\hspace{-5pt}\bar n}\gamma^\rho\slash{\hspace{-5pt}n}\gamma_\nu \omega_\rho^{\ \rho'} 
= \gamma_0\hat H_{\nu\mu}^{(1)\rho\dag}(x_2)\gamma_0\omega_\rho^{\ \rho'}, 
\end{eqnarray}
is either independent of $x_2$ or independent of $x_1$. 
We put them into Eq.(\ref{eq:Wsi2}) and see that, 
for the contribution from the diagram with left cut ($c=L$), 
the only $k_2$-dependence in the integrand is from the matrix element $\hat\Phi^{(1)}_\rho(k_1,k_2,p,S)$, 
while for the contribution from the right cut ($c=R$) diagram 
the only $k_1$ dependence is from the matrix element $\hat\Phi^{(1)}_\rho(k_1,k_2,p,S)$.
Hence, we can carry out the integration over $d^4k_2$ or $d^4k_1$ 
in $\hat\Phi^{(1)}_\rho(k_1,k_2,p,S)$ and obtain, for the 
$\mu\leftrightarrow\nu$ symmetric part, 
\begin{equation}
\tilde W_{\mu\nu}^{(1,si)}(q,p,S,k')=\frac{1}{\pi}{\rm Re}
\int\frac{d^4k}{(2\pi)^4} K(k',k,q)
{\rm Tr}[\hat H_{\mu\nu}^{(1)\rho}(x)\omega_\rho^{\ \rho'}\hat\varphi^{(1)}_{\rho'}(k,p,S)], 
\end{equation}
where $\hat\varphi^{(1)}_\rho(k,p,S)$ is defined as,
\begin{equation}
\hat\varphi^{(1)}_\rho(k,p,S)\equiv
\int \frac{d^4k_2}{(2\pi)^4}\hat\Phi^{(1)}_\rho(k,k_2,p,S)
=\int d^4ze^{ikz}\langle p,S|\bar\psi(0) {\cal L}(0,z)
 D_\rho(z)\psi(z)|p,S\rangle,
\label{eq:corr2}
\end{equation}
The matrix element $\hat\varphi^{(1)}_\rho(k,p,S)$ 
depends only on
one external parton momentum thus is much simpler than the
unintegrated counterpart $\hat\Phi^{(1)}_\rho(k_1,k_2,p,S)$.
This means that, in SIDIS, only $\hat\varphi^{(1)}_\rho(k,p,S)$ 
but not $\hat\Phi^{(1)}_\rho(k_1,k_2,p,S)$ is relevant.

To demonstrate the usefulness of the formalism,
we now calculate the azimuthal asymmetries to the order of $1/Q$ in
SIDIS with unpolarized electron and transversely polarized proton. 
In this case, we need to consider only the 
symmetric parts of $W_{\mu\nu}$ and should  
include the contributions from $\hat W^{(1,si)}_{\mu\nu}$. 
The calculations are in principle straightforward but a little bit involved. 
We need to first decompose the matrix elements in terms of the $\Gamma$-matrices.
Since the hard parts contain only odd-number of $\gamma$'s,  
only the $\gamma_\alpha$ and $\gamma_5\gamma_\alpha$ terms contribute. 
We carry out the traces of $\gamma_\alpha$ and $\gamma_5\gamma_\alpha$ with the 
hard parts, make the Lorentz contraction with the leptonic tensor $L_{\mu\nu}$ 
and obtain the differential cross section as,
\begin{eqnarray}
&&d\sigma=\frac{2\alpha_{em}^2e_q^2}{Q^4}
\frac{dx_BdQ^2d^3k'}{(2\pi)^32E_{k'}}
\int\frac{d^4k}{(2\pi)^4}K(k,k',q)\delta(x-x_B) 
\bigl\{ \sigma^{(0)\alpha}(x)\Phi^{(0)}_\alpha(k,p,S)+ \nonumber\\
&&\phantom{XXXXXXXXXXXXXX}
2{\rm Re}\bigl[\sigma^{(1)\alpha\rho}(x)\varphi^{(1)}_{\alpha\rho}(k,p,S)+
\tilde\sigma^{(1)\alpha\rho}(x)\tilde\varphi^{(1)}_{\alpha\rho}(k,p,S)\bigr]\bigr\}; 
\end{eqnarray}
where $\Phi^{(0)}_\alpha(k,p,S)={\rm Tr}[\gamma_\alpha\hat\Phi^{(0)}(k,p,S)]/2$, 
$\varphi^{(1)}_{\alpha\rho}(k,p,S)={\rm Tr}[\gamma_\alpha\hat\varphi^{(1)}_\rho(k,p,S)]/2$, 
$\tilde\varphi^{(1)}_{\alpha\rho}(k,p,S)=
{\rm Tr}[\gamma_5\gamma_\alpha\hat\varphi^{(1)}_\rho(k,p,S)]/2$;  
and $\sigma^{(0)}(x)$, $\sigma^{(1)\alpha\rho}(x)$ and $\tilde\sigma^{(1)\alpha\rho}(x)$ stand for, 
\begin{eqnarray}
&&\sigma^{(0)\alpha}(x)
=A(y)q^-n^\alpha+ 
2(1-y)x_Bp^\alpha+(2-y)yl_\perp^\alpha, \\
&&\sigma^{(1)\alpha\rho}(x)
=-y^2(l+l')^\alpha l_\perp^\rho/Q^2, \hspace{1cm}
\tilde\sigma^{(1)\alpha\rho}(x)
=-iy^2(l+l')^\alpha |\vec l_\perp| n_{\perp2}^{\rho}/Q^2,
\end{eqnarray}
where $A(y)=1+(1-y)^2$. 

The Lorentz structure of $\Phi_\alpha^{(0)}(k,p,S)$ 
is given by \cite{Mulders:1995dh,ftCorr}, 
\begin{equation}
\Phi_\alpha^{(0)}(k,p,S)
=p_\alpha f_1
+\omega_\alpha^{\ \alpha'}k_{\alpha'}f_\perp
+\varepsilon_{\alpha\beta\gamma\delta}p^\beta k^\gamma S^\delta f_{1T}^{\perp}/M,
\end{equation}
where $M$ is the nucleon mass,  
the $f$'s on the r.h.s. of the equations are 
Lorentz scalars and are functions of $k\cdot p$ and $k^2$.
Note that by integrating $\Phi_\alpha^{(0)}$
over $k^-$, we obtain the $k_\perp$-dependent 
quark distributions $\Phi_\alpha^{(0)}(x,\vec k_\perp)$ 
and that $\int p^+dk^- f_{1T}^\perp$ is just 
the Sivers function\cite{Sivers:1989cc} 
discussed in connection with single-spin asymmetries. 

There are much more terms involved for the Lorentz structure of 
$\varphi^{(1)}_{\alpha\rho}$ and 
$\tilde\varphi^{(1)}_{\alpha\rho}$. 
We made a complete examination of them and found out that, 
to the order of $1/Q$, the contributing terms are,
\begin{eqnarray}
\varphi^{(1)}_{\rho\alpha}(k,p,S)
&=& k_\rho p_\alpha\varphi^{(1)}_\perp+
Mp_\alpha\varepsilon_{\perp\rho\delta}S^\delta\varphi^{(1)}_{\perp s1}
+p_\alpha \varepsilon_{\perp\gamma\delta} 
(k_\rho k^\gamma -k_\perp^2g_\rho^{\ \gamma}) 
S^\delta\varphi^{(1)}_{\perp s2}/M+..., \\
\tilde\varphi^{(1)}_{\rho\alpha}(k,p,S)
&=&ip_\alpha\varepsilon_{\perp\rho\gamma}k^\gamma \tilde\varphi_\perp^{(1)} +
iMp_\alpha S_\rho\tilde\varphi_{\perp s1}^{(1)}
+ ip_\alpha 
\varepsilon_{\perp\rho\beta}\varepsilon_{\perp\gamma\delta}
(k^\beta k^\gamma -k_\perp^2g^{\beta\gamma}) 
S^\delta\tilde\varphi_{\perp s2}^{(1)}/M+..., 
\end{eqnarray}
where $\varepsilon_{\perp\rho\alpha}=\varepsilon_{\rho\alpha\gamma\delta}\bar n^\gamma n^\delta$, 
and all the $\varphi$'s and $\tilde\varphi$'s on the r.h.s. of 
the equations are Lorentz scalars and functions of $k\cdot p$ and $k^2$.

Equation of motion relates the $\varphi$'s and $\tilde\varphi$'s to $f$'s as, 
\begin{equation}
xf_\perp=-\varphi^{(1)}_{\perp}+\tilde\varphi^{(1)}_{\perp}, \hspace{2cm} 
xp\cdot kf^\perp_{1T}=-M^2(\varphi^{(1)}_{\perp s1}+\tilde\varphi^{(1)}_{\perp s1}).
\end{equation}

Using the above expansion of the parton matrix elements, 
one obtains that, to the order $1/Q$, 
\begin{eqnarray}
\label{eq:csres}
d\sigma&=&\frac{2\alpha_{em}^2e_q^2}{Q^4}
\frac{dx_BdQ^2p^+dk^-d^2k_\perp}{(2\pi)^4} 
\Bigl\{ A(y)\bigl[
f_1+\frac{|\vec k_\perp|}{M}f_{1T}^\perp \sin(\phi-\phi_s) \bigr] \nonumber\\ 
&&-\frac{MB(y)}{Q}
\bigl[ \frac{2|\vec k_\perp|}{M}(x_Bf_\perp-\tilde\varphi^{(1)}_\perp) \cos\phi 
+\frac{\vec{k}_\perp^2}{M^2}(\varphi_{\perp s2}^{(1)}-2\tilde\varphi_{\perp s2}^{(1)})\sin\phi_s  
+\frac{\vec{k}_\perp^2}{M^2}\tilde\varphi_{\perp s2}^{(1)}\sin(2\phi-\phi_s)\bigr]\Bigr\}, 
\end{eqnarray}
where $B(y)=2(2-y)\sqrt{1-y}$, 
$\phi$ and $\phi_s$ are respectively the azimuthal angle of 
$\vec k_\perp$ and $\vec S_\perp$
relative to the lepton plane;
$\vec k_\perp=\vec {k'}_\perp$, $k^-={k'}^--q^-$ and $k^+=x_Bp^+$.

From Eq.(\ref{eq:csres}), one notes that the leading contribution to 
the azimuthal angle dependence comes only from the leading twist
quark distribution $\Phi_\alpha^{(0)}$ while the higher order ($1/Q$)
receives contribution from both the leading and the next leading
twist parton matrix elements.
In view that the energies of the current polarized experiments, 
in particular at HERMES \cite{Airapetian:2004tw}, are not very high, 
these $1/Q$-terms can be very important. 

The $\sin\phi_s$-terms in Eq.(\ref{eq:csres}) warrants a
special examination since they correspond to single-spin asymmetry. 
It can be seen more clearly from the cross section integrated over $\phi$, i.e,  
\begin{equation}
d\sigma=\frac{\alpha_{em}^2e_q^2}{Q^4}
\frac{dx_BdQ^2p^+dk^-d{\vec{k}_\perp}^2}{(2\pi)^3}
\Bigl\{A(y)f_1 -\frac{B(y)\vec{k}_\perp^2}{MQ}(\varphi_{\perp s2}^{(1)}-2\tilde\varphi_{\perp s2}^{(1)}) 
\sin\phi_s  \Bigr\}. 
\label{eq:cs2}
\end{equation}
\end{widetext}
The single-spin asymmetry $A_N$ is defined as the difference between 
the cross section at $\phi_s=\pi/2$ and that at $\phi_s=3\pi/2$ 
divided by their sum. We see that there exists a finite $A_N$ 
for SIDIS at a given $k^-$ at the order $M/Q$.
This corresponds to one of the asymmetries discussed in \cite{Boer:1997nt}.
Such an asymmetry can provide important information on the
unintegrated parton correlations at given $k^-$ and $k_\perp$. 
The value of $k^-$ could be determined from $k^-={k'}^--q^-$ 
by measuring the momenta of the scattered lepton and 
single jet.
This could be done with reasonable accuracy at very high energies
where one can reconstruct the single jet event.   
However it might be very difficult at lower energies.

If one integrates over $k^-$ and $\vec k_\perp$ 
to obtain the 
inclusive cross section for $e^-p\to e^-X$, 
these $\sin\phi_s$-terms should vanish as demanded by 
the parity and time reversal invariance
of the inclusive hadronic tensor. 
This implies that the integration of 
$\varphi_{\perp s}^{(1)}$ and $\tilde\varphi_{\perp s}^{(1)}$  
over $k^-$ and $k_\perp$ should vanish for unrestricted ranges of $k^-$ and $k_\perp$, 
leading to zero single-spin asymmetry for inclusive DIS. 
However, if we integrate over $k^-$ only for a restricted range in which
the outgoing partons are time-like, {\it i.e.} for $0<{k'}^-<\infty$, 
the result might be non-zero. 
In practice, this is equivalent to 
identifying a large momentum final hadron 
in the photon direction to guarantee that the outgoing parton is time-like.
We call such events as ``triggered inclusive process'' and denotes
it by $e^-p\to e^- + h_{trig}+X$. 
The averaged single-spin asymmetry
could be finite for such triggered inclusive DIS.

At the hadron level, such a single-spin asymmetry 
corresponds to a spin-dependent term 
in the hadronic tensor $W_{\mu\nu}^{(trig)}(q,p,S)$, i.e., 
\begin{eqnarray}
&&\frac{1}{2M}W_{\mu\nu}^{(S,trig)}(q,p,S)=\frac{1}{2p\cdot q}
[\varepsilon_{\perp\mu\gamma}(q_\nu+2x_Bp_\nu)+  \nonumber\\
&&\phantom{XXX}
\varepsilon_{\perp\nu\gamma}(q_\mu+2x_Bp_\mu)]S^\gamma G_s(x_B,Q^2).
\end{eqnarray}
and the single-spin spin asymmetry $A_N^{(trig)}$ is given by, 
\begin{equation}
A_N^{(trig)}= \frac{d\sigma^\uparrow-d\sigma^\downarrow}
         {d\sigma^\uparrow+d\sigma^\downarrow}
=B(y)
\frac{x_BM}{Q}\frac{G_s}{F_1},
\end{equation}
where $F_1$ is the normal spin averaged structure function.
In terms of the parton correlations discussed above, 
$G_s$ is given by, 
\begin{equation}
x_BG_s
=-\int\frac{d^4k}{(2\pi)^4}\delta(x-x_B)
\frac{\vec{k}_\perp^2}{M^2}(\varphi_{\perp s2}^{(1)}-2\tilde\varphi_{\perp s2}^{(1)}).
\end{equation}
The integration limit for $k^-$ is $-q^-<k^-<\infty$.
Experimental measurements of the above averaged single-spin asymmetry
in the triggered DIS would lead to useful information 
on the spin structure of nucleon.

If we neglect the final-state soft gluon interactions, {\it i.e.} set $g=0$, 
the covariant derivatives in Eq.~(\ref{eq:corr2}) become normal derivatives. 
We then have $\hat\varphi^{(1)}_\rho=-k_\rho\hat\phi^{(0)}$, 
and $-\varphi^{(1)}_\perp=f_1=xf_\perp$, 
$\varphi^{(1)}_{\perp s1}=\varphi^{(1)}_{\perp s2}=0$,
$\tilde\varphi^{(1)}_{\perp}=\tilde\varphi^{(1)}_{\perp s1}
=\tilde\varphi^{(1)}_{\perp s2}=0$. 
Hence, only $A(y)f_1+2B(y)(|\vec k_\perp|/Q)f_1\cos\phi$ are left
in Eq.(\ref{eq:csres}) 
which is the result in Ref.~\cite{Cahn:1978se}.\cite{ftBM}
The difference between this and the full result given by Eq.(\ref{eq:csres}) 
comes from the final-state soft gluon interactions.
Data on $\langle\cos\phi\rangle$ obtained in unpolarized 
experiments\cite{Aubert:1983cz,Arneodo:1986cf,Adams:1993hs,Breitweg:2000qh,Chekanov:2002sz}
suggest the existence of such QCD contributions. 
It is also obvious that fragmentation can contribute to 
the azimuthal asymmetries discussed above. 
A complete formalism for SIDIS should include fragmentation. 
Such an extension is underway.


\section{Summary}

In summary, we derived the factorized form 
of the cross section for 
semi-inclusive deep-inelastic lepton-nucleon scattering 
as a convolution of the hard parts with the 
gauge invariant unintegrated parton distributions 
and higher twist parton correlations. 
As a consequence of the collinear expansions,  
the hard parts depend only on the 
longitudinal components of the parton momenta.
The gauge invariant parton distributions receive 
contributions from initial and final state interactions 
and provide the only dependence on the initial parton transverse momentum. 
Results for azimuthal angle dependence to the order of $1/Q$ 
in reactions with transversely polarized targets are given.
A novel single-spin asymmetry for the ``triggered inclusive 
process'' $e^-p^\uparrow\to e^-+h_{trig}+X$ is predicted to 
the order of $M/Q$.

We thank X.~D. Ji, J.~P. Ma and J. Zhou for helpful discussions. 
This work was supported in part by 
the U.S. Department of Energy 
under No. DE-AC03-76SF00098; 
and National Natural Science Foundation of China
under the approval Nos. 10525523 and 10440420018.


\vspace{-0.2in}
\begin{thebibliography}{99}

\bibitem{Aubert:1983cz}
  J.~J.~Aubert {\it et al.}  [European Muon Collaboration],
  Phys.\ Lett.\ B {\bf 130}, 118 (1983);
%
\bibitem{Arneodo:1986cf}
  M.~Arneodo {\it et al.}  [European Muon Collaboration],
  Z.\ Phys.\ C {\bf 34}, 277 (1987).

\bibitem{Adams:1993hs}
  M.~R.~Adams {\it et al.}  [E665 Collaboration],
  Phys.\ Rev.\ D {\bf 48}, 5057 (1993).

\bibitem{Breitweg:2000qh}
  J.~Breitweg {\it et al.}  [ZEUS Collaboration],
  Phys.\ Lett.\ B {\bf 481}, 199 (2000);
%
\bibitem{Chekanov:2002sz}
  S.~Chekanov {\it et al.}  [ZEUS Collaboration],
  Phys.\ Lett.\ B {\bf 551}, 226 (2003).
%
\bibitem{Airapetian:2004tw}
  A.~Airapetian {\it et al.}  [HERMES Collaboration],
  %
  Phys.\ Rev.\ Lett.\  {\bf 94}, 012002 (2005).

\bibitem{Alexakhin:2005iw}
  V.~Y.~Alexakhin {\it et al.}  [COMPASS Collaboration],
  Phys.\ Rev.\ Lett.\  {\bf 94}, 202002 (2005);
%
\bibitem{Webb:2005cd}
  R.~Webb  [COMPASS Collaboration],
  %
  Nucl.\ Phys.\ A {\bf 755}, 329 (2005).

\bibitem{Georgi:1977tv}
  H.~Georgi and H.~Politzer,
  Phys.\ Rev.\ Lett.\  {\bf 40}, 3 (1978).

\bibitem{Cahn:1978se}
  R.~N.~Cahn,
  Phys.\ Lett.\ B {\bf 78}, 269 (1978).

\bibitem{Berger:1979kz}
  E.~L.~Berger,
  Phys.\ Lett.\ B {\bf 89}, 241 (1980).

\bibitem{Liang:1993re}
  Z.~T.~Liang and B.~Nolte-Pautz,
  Z.\ Phys.\ C {\bf 57}, 527 (1993).

\bibitem{Oganesian:1997jq}
  K.~A.~Oganesian, H.~R.~Avakian, N.~Bianchi and P.~Di Nezza,
  Eur.\ Phys.\ J.\ C {\bf 5}, 681 (1998).

\bibitem{Chay:1997qy}
  J.~Chay and S.~M.~Kim,
  Phys.\ Rev.\ D {\bf 57}, 224 (1998).

\bibitem{Collins:1992kk}
J.~C.~Collins,
Nucl.\ Phys.\ B {\bf 396}, 161 (1993);
  Phys.\ Lett.\ B {\bf 536}, 43 (2002).



\bibitem{Mulders:1995dh}
  P.~J.~Mulders and R.~D.~Tangerman,
  Nucl.\ Phys.\ B {\bf 461}, 197 (1996)
  [Erratum {\bf 484}, 538 (1997)].

\bibitem{Boer:1997nt}
  D.~Boer and P.~J.~Mulders,
  Phys.\ Rev.\ D {\bf 57}, 5780 (1998).

\bibitem{Boer:1999uu}
  D.~Boer, R.~Jakob and P.~J.~Mulders,
  Nucl.Phys. B{\bf 564}, 471 (2000).

\bibitem{Ji:2002aa}
  X.~Ji and F.~Yuan,
  Phys.\ Lett.\ B{\bf 543}, 66(2002).

\bibitem{Belitsky:2002sm}
  A.~Belitsky, X.~Ji and F.~Yuan,
  Nucl.\ Phys.\ B{\bf 656}, 165(2003).

\bibitem{Ji:2004wu}
  X.~D.~Ji, J.~P.~Ma and F.~Yuan,
  Phys.\ Rev.\ D {\bf 71}, 034005 (2005).
%
\bibitem{Ji:2005nu}
  X.~D.~Ji, J.~P.~Ma and F.~Yuan,
  JHEP {\bf 0507}, 020 (2005).

\bibitem{Anselmino:2005ea}
  See e.g. talks at the 
  international workshop on
  transverse polarization phenomena in
  hard processes (Transversity 2005), 
  Villa Olmo, September 2005, e.g.,  
  M.~Anselmino {\it et al.}, 
  hep-ph/0507181.

\bibitem{Ellis:1982wd}
  R.~K.~Ellis, W.~Furmanski and R.~Petronzio,
%
  Nucl.\ Phys.\ B {\bf 207}, 1 (1982); {\bf 212}, 29 (1983).
%
  %


\bibitem{Qiu:1990xx}
  J.~W.~Qiu and G.~Sterman,
  Nucl.\ Phys.\ B {\bf 353}, 105 (1991); {\bf 353}, 137 (1991).
%

\bibitem{Guo:2000nz}
  X.~F.~Guo and X.~N.~Wang,
  Phys.\ Rev.\ Lett.\  {\bf 85}, 3591 (2000);
  X.~N.~Wang and X.~F.~Guo,
  Nucl.\ Phys.\ A {\bf 696}, 788 (2001).

\bibitem{Sivers:1989cc}
  D.~W.~Sivers,
  Phys.\ Rev.\ D {\bf 41}, 83 (1990); {\bf 43}, 261 (1991).

\bibitem{ftQiu}
The gauge link presented here makes the parton correlations 
gauge invariant only when $A^-(z^+,z^-,\vec z_\perp)\to 0$ when $z^-\to\infty$.
This is not valid in the light gauge. 
In that case, special treatment is needed. 
For a more general form of ${\cal{L}}$, see J.W. Qiu, paper in preparation.
We thank Jian-Wei for discussion in this and other related connections.    

\bibitem{ftCorr} We note that, 
$\omega_\alpha^{\ \alpha'}k_{\alpha'}f_\perp= 
k^-n_{\alpha'}f_\perp+k_{\perp\alpha}f_\perp$, 
this corresponds to two independent terms after integration over $k^-$. 
Similarly, 
$\vec k_\perp^2\varepsilon_{\alpha\beta\gamma\delta}p^\beta k^\gamma S^\delta 
=(p\cdot k)[(k_\perp\cdot S_\perp)\varepsilon_{\perp\alpha\gamma}k_\perp^\gamma 
+k_{\perp\alpha}\varepsilon_{\perp\gamma\delta}k_\perp^\gamma S_\perp^\delta]
-\vec k_\perp^2p_\alpha \varepsilon_{\perp\gamma\delta}k_\perp^\gamma S_\perp^\delta$, 
we obtain also more than one terms from 
$\varepsilon_{\alpha\beta\gamma\delta}p^\beta k^\gamma S^\delta f_{1T}^\perp$ 
after integrating over $k^-$. 
That is why we have even more independent terms for the correlators after the 
integration over $k^-$. See e.g. 
  K.~Goeke, A.~Metz and M.~Schlegel,
  Phys.\ Lett.\  B {\bf 618}, 90 (2005)
  [arXiv:hep-ph/0504130].


\bibitem{ftBM} We note that, in [16], 
the authors started with an expression where the hard part 
contains transverse components which lead to double-counting 
of the transverse contributions. 
But they have either the projection operator  
$\omega_\rho^{\ \rho'}$ or the $\partial_\rho$-term in $\varphi^{(1)}$. 
This is why they came also to this result at $g=0$. 

\end{thebibliography}
\end{document}